Limitations of P-Values and $R^2$ for Stepwise Regression Building:

A Fairness Demonstration in Health Policy Risk Adjustment

Sherri Rose and Thomas G. McGuire

Department of Health Care Policy, Harvard Medical School, Boston, MA

Sherri Rose is Associate Professor, Department of Health Care Policy, Harvard Medical School, Boston, MA, 01201 (E-mail: rose@hcp.med.harvard.edu). Thomas McGuire is Professor, Department of Health Care Policy, Harvard Medical School, Boston, MA 01201 and Research Associate at NBER, Cambridge, MA, 02138. This work was supported by NIH grant DP2-MD012722 and the Laura and John Arnold Foundation. The authors thank Monica Farid for data preparation.

**Abstract**

Stepwise regression building procedures are commonly used applied statistical tools, despite their well-known drawbacks. While many of their limitations have been widely discussed in the literature, other aspects of the use of individual statistical fit measures, especially in high-dimensional stepwise regression settings, have not. Giving primacy to individual fit, as is done with p-values and $R^2$, when group fit may be the larger concern, can lead to misguided decision making. One of the most consequential uses of stepwise regression is in health care, where these tools allocate hundreds of billions of dollars to health plans enrolling individuals with different predicted health care costs. The main goal of this "risk adjustment" system is to convey incentives to health plans such that they provide health care services fairly, a component of which is not to discriminate in access or care for persons or groups likely to be expensive. We address some specific limitations of p-values and $R^2$ for high-dimensional stepwise regression in this policy problem through an illustrated example by additionally considering a group-level fairness metric.

1. INTRODUCTION

Many statistics textbooks, including those in graduate-level statistical science and applied health courses, instruct students to perform stepwise regressions for practical data analyses. These methods have been deployed in the context of building prediction functions, where input variables are used to estimate outcomes, and effect estimation, where the parameter of interest is instead the effect of an exposure or intervention on the outcome. Stepwise procedures have varied flavors and may start with an empty or partial model adding variables, a full model with all variables followed by deletion steps, or some combination or these two processes (Efroymson 1966; Draper and Smith 1966; Mantel 1970; Hocking 1976; Miller 2002; Sinisi and van der Laan 2004). After each variable addition or deletion step, an evaluation metric is used to assess whether that step impacted fit. Metrics can include p-values, $R^2$, F tests, and additional choices depending on the procedure.

However, there are numerous widely discussed limitations of stepwise methods, such as misleadingly small p-values not adjusted to account for the iterative fitting and biased $R^2$ measures, among others (Hocking 1976; Harrell 2001). Post-selection inference is an active area of statistical methods research, but most off-the-shelf statistical software does not incorporate these advances. Despite cautioning applied researchers about the shortcomings of traditional stepwise procedures, they are commonly used in practice.

One empirical prediction setting where stepwise regressions are currently applied is risk adjustment in health policy, which spans many areas, including public reporting, quality measurement, and payment models for health insurance plans. This paper focuses on the problem of stepwise regression building for the latter case, the prediction of payments to health plans, as a demonstrative example. We summarize how stepwise regression works in the context of plan payment risk adjustment to predict individual spending as well as some central limitations of p-values and $R^2$ for high-dimensional prediction. We then consider aspects of "fairness" and implement a metric to assess fair predictions for individuals in vulnerable (i.e., undercompensated) groups. Groups, such as individuals with mental health and substance use disorders (MHSUD), are considered according to the means a health plan has at their disposal to discourage enrollment in their health plans and restrict health services.

2. PROBLEMS WITH P-VALUES AND $R^2$ IN STEPWISE REGRESSION FOR RISK ADJUSTMENT

Risk adjustment in health plan payment as translated into a statistical estimation framework is a familiar, classical prediction question, which we define formally here. Consider a continuous spending outcome $Y$, a vector of demographic variables $\boldsymbol{D} = (D_1, D_2, \ldots, D_K)$, and a vector of health variables $\boldsymbol{H} = (H_1, H_2, \ldots, H_J)$. A typical risk adjustment model would be given by:

$$E(Y|\boldsymbol{D}, \boldsymbol{H}) = \alpha \boldsymbol{D} + \beta \boldsymbol{H},$$

where $E$ represents an expectation, $E(Y|\boldsymbol{D}, \boldsymbol{H})$ is the condition mean of $Y$ given $(\boldsymbol{D}, \boldsymbol{H})$, $\alpha$ are coefficients for $\boldsymbol{D}$ demographic variables, and $\beta$ are coefficients for the $\boldsymbol{H}$ health variables. The goal is the estimate $E(Y|\boldsymbol{D}, \boldsymbol{H})$, our regression function, in order to predict spending. These predictions feed into the determination of the payment the health plan receives in exchange for accepting responsibility for paying for the individual's health care costs. Reasonably accurate prediction is necessary to ensure that health plans, many of which are for-profit, will be willing to accept and provide good care to enrollees with high costs.

Demographic variables generally take the form of age-sex cell indicator variables. Most of the effort in the development of plan payment risk adjustment formulas is associated with the creation and selection of the health variables. Thus, as a simplified introductory example, stepwise regression in this setting could take the following form:

1. Start with a baseline formula containing only demographic variables $\boldsymbol{D}$: $E(Y|\boldsymbol{D}) = \alpha\boldsymbol{D}$.

2. Assign a rule that will keep added health variables $\boldsymbol{H} = (H_1, H_2, \ldots, H_J)$ in the formula if they have significant p-values and also increase $R^2$.

3. Sequentially add health variables $\boldsymbol{H}$, beginning with $H_1$, assessing p-values and $R^2$ metrics after each addition.

This illustration could be criticized for many established reasons, including the choice to introduce variables one at a time, that it will not evaluate all possible combinations of variables, and that we may not end up with an optimal or near-optimal selection of variables in the final formula (Mantel 1970; Hocking 1976). We choose to focus on the issue of fairness metrics vs. p-values and $R^2$ as yet another consideration.

Although researchers and regulators recognize a number of criteria for inclusion of variables in risk adjustment formulas, in practice, the most influential statistical criterion appears to be whether a given variable improves performance, as measured by p-value or $R^2$ statistic. These metrics are automatically produced by statistical programs (oftentimes without appropriate correction for the repeated model fitting), but no metric is automatically generated to capture performance in other, potentially more important dimensions. Simply put, using a p-value or $R^2$ as the statistical criterion to decide variable inclusion in the prediction function can lead to mistakes, in the sense that improving fit at the person level may not improve fit in the ways that may have even more impact, at the group level.

We do not reiterate many of the arguments on the disadvantages of p-values, for example, as enumerated in Wasserstein and Lazar (2016). However, before defining our group-level fit metric to evaluate fairness, we do highlight two features of p-values that are particularly relevant for stepwise regression building in risk adjustment, which is a high-dimensional prediction problem. Firstly, as is well-known, large sample sizes will yield significant p-values for variable coefficients that are of trivial magnitude (e.g., Chatfield 1995; van der Laan and Rose 2010). Many plan payment risk adjustment formulas are created using millions of observations; assessing a variable's importance for predicting health spending using p-values is effectively useless. The second is regarding the relationship between $R^2$ and p-values. We keep in mind that substantial improvements in $R^2$ are accompanied by significant p-values for the added variable, but also that miniscule improvements in $R^2$ may result from the addition of a variable with a nonsignificant *or* significant p-value.

## 3. ADDITIONAL METRICS FOR RISK ADJUSTMENT

The most important criteria for evaluating a risk adjustment scheme follow from the efficiency or fairness problem risk adjustment is trying to fix (Layton et al. 2017). In the U.S., Medicare Advantage (for older and some disabled adults), Medicaid Managed Care (for people with lower incomes) and the Marketplaces (for otherwise uninsured, created as part of the Affordable Care Act) all operate as individual health insurance markets with competing health plans. In these systems, health plans are prohibited by law from discriminating in enrollment or services against *individuals*. For example, plans must accept any individual who applies for membership. However, plans can and do discriminate against *groups* of individuals, such as those with MHSUD, by: (*a*) limiting provider networks treating this disorder, (*b*) setting low provider payments to mental health providers to discourage supply, (*c*) providing less favorable coverage of drugs, and other means. Consequently, individual-level fit is secondary to group-level fit as a metric for alternative plan payment schemes.

Group-level fit in U.S. plan payment risk adjustment formulas is often measured by predictive ratios, equal to the ratio of predicted values over actual values for a group. A predictive ratio less than one indicates that the prediction function underpredicts and will therefore underpay for the group. Layton et al. (2017) found that underpayment as measured by the predictive ratio in the Marketplaces was most severe for the mental illness group, among the four groups they examined. This is concerning because, as noted, although a Marketplace plan must accept all applicants, the plan can provide poor care for MHSUD, discouraging individuals with these conditions from seeking enrollment in the first place.

In Europe, it is more common to measure group fit by net compensation, equal to the average difference between predicted values for a particular group and actual values:

$$\text{Net Compensation} = \frac{\sum_{i \in g} \hat{Y}_i}{n_g} - \frac{\sum_{i \in g} Y_i}{n_g},$$

where $\hat{Y}_i$ is predicted spending for individual *i*, $Y_i$ is observed spending for individual *i*, and $n_g$ is the sample size for the group of interest (Layton et al. 2017). The sums are taken over all individuals in the group. Net compensation measures incentives to a plan to provide good service to a group. Groups need not be mutually exclusive, and net compensation can be defined with respect to each health condition group. We use this fairness metric in our demonstration as it is on the same scale as our outcome of interest, and therefore has an easy interpretation.

Other metrics of health plan performance have been implemented or proposed in the case of ensuring equal access to mental health care, but these are recognized as incomplete and of doubtful effectiveness as a basis for monitoring plan services (McGuire 2016). While many fairness metrics focus on classification problems, individual vs. group-based notions of fairness have been studied (e.g., Zemel et al. 2013; Hu and Chen 2017), as well as general frameworks that include non-classification problems (e.g., Kusner et al. 2017). We refer to Mitchell (2017) for a didactic summary of fairness metrics in the machine learning and computer science literature.

## 4. DEMONSTRATION OF THE METRICS

We include an instructive example using the Truven Marketscan data for plan payment risk adjustment, which mirrors the approach implemented in the Marketplaces with some modification. Plan payment in the Marketplaces is complex, involving adjustments for geographic factors and premiums the plans collect from enrollees, among other factors (Layton et al. 2018). At the core of the payment scheme, however, is the Department of Health and Human Services Hierarchical Condition Category (HSS-HCC) linear least squares regression prediction model, which determines the base payment for each individual (Centers for Medicare and Medicaid Services 2016).

Each HCC variable is the result of a mapping from a subset of the thousands of five-digit International Classification of Disease and Related Health Problems (ICD) diagnoses reported on claims to a much smaller number of categories. For example, the HCC for "major depressive, bipolar, and paranoid disorders" is generated from over 50 ICD-9 flags. It is important to note that not all ICD-9 (or ICD-10, adopted in 2015) codes map to an HCC used for payment. This has been shown to be problematic with respect to accurate payments, especially for MHSUD in the Marketplaces (Montz et al. 2016). Montz et al. (2016) found systematic underpayment of enrollees with MHSUD; 80% of individuals with MHSUD are not recognized by the Marketplace system, contributing to undercompensation for individuals with MHSUD on average.

The HHS-HCC model undergoes regular evaluation, as does its progenitor, the Centers for Medicare and Medicaid (CMS) version used for Medicare Advantage, including consideration of the HCC diagnostic adjustors. The HCCs used in the HHS-HCC model are a subset of the full 264 HCCs defined in the full system mapping. Adding or subtracting HCCs from the right-hand side of the risk adjustment formula is a component in the evaluation of the prediction function. Government reports lay out the criteria used in defining HCCs, with the first two being the HCC should be "clinically meaningful" and "predictive" (Pope et al. 2004; Ellis et al. 2018). Other risk adjustment formulas consider additional variable types for special populations, such as measures of functional status in the CMS frailty model (Kautter and Pope 2004).

Our Marketscan sample contained all those continuously enrolled from January 2015 to December 2016 who had prescription drug coverage and mental health coverage. We excluded enrollees with missing geographic region or claims information, as well as those with negative claims. With these restrictions, we took a sample of 4,000,000 enrollees that we used in our analysis. The outcome, total annual expenditures in 2016, was calculated by summing all inpatient, outpatient, and drug payments. Mean total spending for an individual adult was $6,619.

The baseline formula included 75 HCCs from 2015, as well as age and sex, as predictor variables in a main terms parametric regression. We additionally considered two mental health HCCs and two substance use disorder HCCs. These 79 HCCs followed the CMS-HCC risk adjustment formula in place for Medicare Advantage (Pope et al. 2011). We defined the MHSUD group for calculation of net compensation using Clinical Classification Software (CCS) categories, a more comprehensive set of variables compared to the HCCs. Each ICD flag maps to a CCS category, unlike the mapping from ICD to HCC described above. Therefore, our calculation of net compensation will capture the impact of the risk adjustment formula for those with MHSUD recognized *and* unrecognized by the formula. (We emphasize here that individuals with MHSUD but no MHSUD ICD flags will not be captured by HCCs or CCS categories.) The MHSUD group contains 13.8% of the sample, compared to the 2.6% of the sample identified using the four HCCs. Mean total spending for an individual in the MHSUD group was $11,346, which was 71% higher than mean spending for an individual in the total sample.

We describe possible iterative stepwise regression decision-making processes in Figure 1, guided by different metrics (that could be defined a priori and within cross-validation, e.g., to try to avoid cherry picking and overfitting). As noted earlier, with a sample size of four million enrollees, p-values are a largely ineffective metric. All but three variables in the baseline formula were significant, even those with small event rates, such as "pressure ulcer of the skin, with necrosis through to muscle, tendon, bone" with 109 enrollees. The most prevalent HCC was "diabetes without complications," occurring among 5.0% of individuals in the sample. This baseline formula had an adjusted $R^2$ of 13.1%, which is similar to the 12% $R^2$ typically achieved with the same specification in Medicare. Net compensation for MHSUD was a nontrivial underpayment, reflected in the negative value -$2,822, which was 25% of the mean total spending in the MHSUD group. Therefore, enrollees with MHSUD are vastly underpaid relative to those without MHSUD, giving insurers a strong incentive to distort their plan offerings to avoid these enrollees.

**Figure 1. Decision-Making Flow Chart for Plan Payment Formula with Differing Metrics**

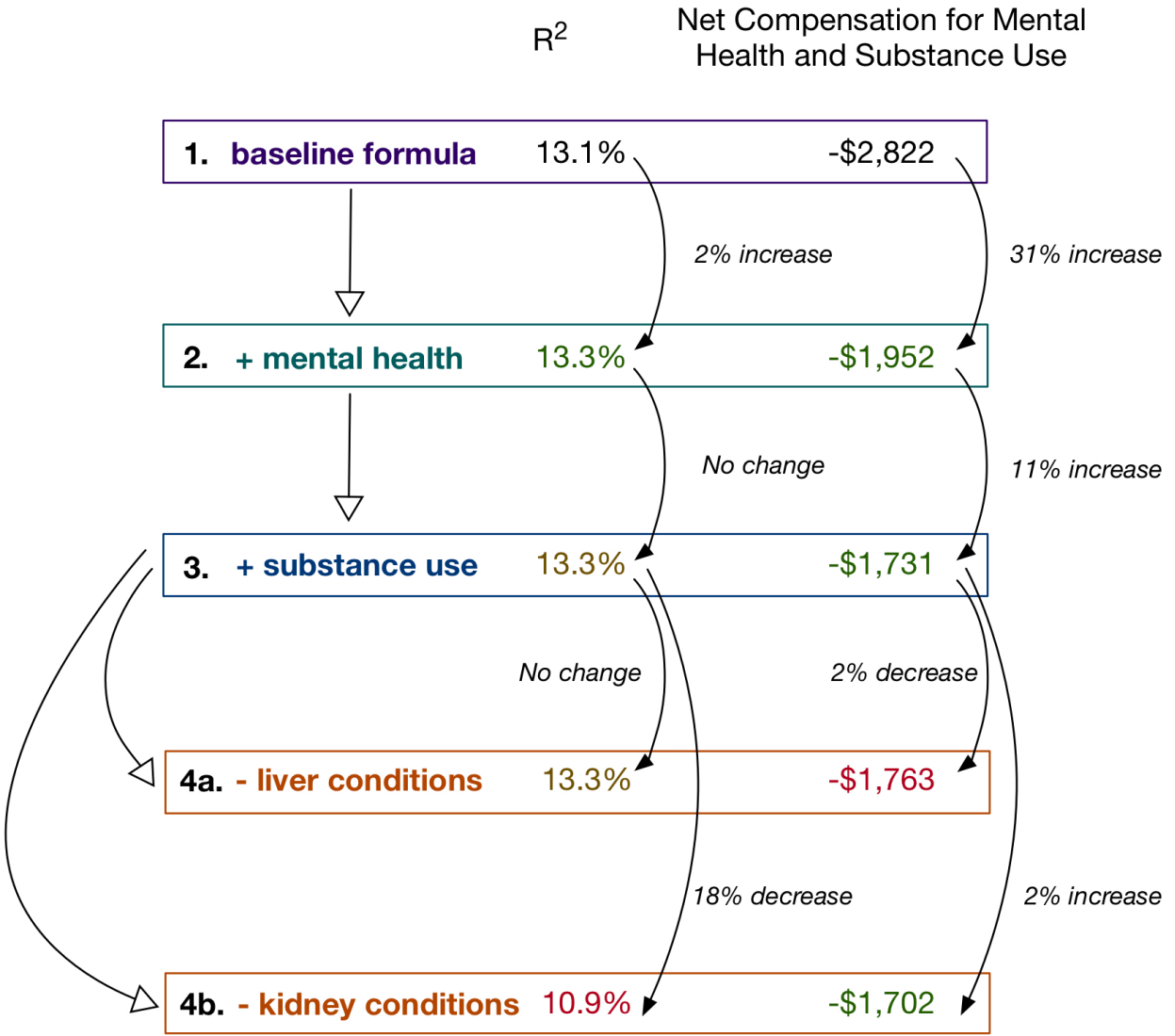

A stepwise regression formula building exercise that aims to maximize $R^2$ *or* optimize MHSUD net compensation toward zero might proceed from the baseline formula (formula 1 in Figure 1) to formula 2, which added two mental health HCCs ("schizophrenia" and "major depressive, bipolar, and paranoid disorders"). This provided a small 2% increase in $R^2$ but a 31% increase in MHSUD net compensation; p-values for both variables were statistically significant. Because MHSUD net compensation calculated based on formula 1 was a large negative number, *increases* in MHSUD net compensation that did not lead to a value greater than zero were improvements. Moving from formula 2 to formula 3, we added two substance use HCCs ("drug/alcohol psychosis" and "drug/alcohol dependence"). The $R^2$ did not change; a stepwise regression maximizing based on $R^2$ (and parsimony) would revert back to formula 2. However, there was a continued increase in MHSUD net compensation (this time of 11%), thus a stepwise regression optimizing this metric would keep formula 3 and continue.

Moving from formula 3, which contained the complete set of 79 HCCs used in the Medicare Advantage risk adjustment formula, the stepwise regression might then consider deletion steps of statistically significant variables to improve fit and fairness. Formula 4a removed the HCCs associated with liver conditions with no impact on $R^2$, but a 2% decrease in MHSUD net compensation (a movement in the wrong direction). A stepwise regression driven by $R^2$ and parsimony would prefer such a formula, despite being worse than formula 3 for MHSUD net compensation (and likely worse for those with liver conditions). An alternative deletion step from formula 3 might be to remove kidney condition HCCs, represented in formula 4b. Here, $R^2$ dropped by 18%, yet MHSUD net compensation improved by 2%, and was indeed the best MHSUD net compensation of all the formulas in Figure 1. A stepwise regression that is searching based solely on improving MHSUD net compensation selects formula 4b, despite its overall poorer global fit (as assessed by $R^2$) and impact on those with kidney conditions.

5. DISCUSSION

Beyond the well-known drawbacks of stepwise regression procedures, building an "appropriate" stepwise risk adjustment formula in a policy environment is not as simple as defining overall fit metrics for evaluation. Overreliance on statistical measures of global fit, such as p-values and $R^2$, in observational individual-level data does not consider the inequalities created or exacerbated in potentially vulnerable groups. Fair formulas are especially critical in healthcare systems where risk adjustment may have a direct or indirect impact on human health. In this context, p-values and $R^2$ may not be "morally neutral," to quote Professor Unsworth's use of this phrase for the subjectivity and ethics of statistical algorithms (Reyes 2016).

The massive size of the health care sector in the U.S. and its salience for personal and social welfare make risk adjustment arguably one of the most consequential applications of stepwise regression for social policy. As another example, valid evaluation of the relative performance of hospitals with respect to quality and outcomes requires risk adjustment for the health of each hospital's patient population, often referred to as "case-mix" (Shahian and Normand 2008). Hospitals with sicker patients along dimensions not included in the risk adjustment formula will perform poorly in these assessments, leading to payment sanctions in public health insurance programs and possible closure by state regulators. Some approaches to case-mix adjustment for hospital quality explicitly use p-values to select variables in more parsimonious models (e.g., O'Malley et al. 2005).

We argue that there is a pressing need to consider a formal ensemble of metrics for evaluation of plan payment risk adjustment that balances both global fit and multiple fairness metrics. This is a likely scenario in other applied settings. It is well-accepted that evaluation of risk adjusted models for plan payment involves numerous criteria. While regulators can and do make good faith efforts to examine vulnerable groups using predictive ratios and other measures (Layton et al. 2016), most of these procedures are ad hoc. Our stepwise regression risk adjustment demonstration represented a simplification of a deeply difficult policy problem and considered only one vulnerable group – those with MHSUD. Even in this simplified example with only one fairness metric, the "best" MHSUD net compensation was still a large underpayment relative to mean total spending, further highlighting the challenge of deploying a more comprehensive system with multiple groups. While unlikely to solve the totality of the fairness issues plaguing plan payment risk adjustment, there are numerous planned changes for the 2019 Medicare Advantage plan payment risk adjustment formula, including adding additional MHSUD HCCs (Centers for Medicare & Medicaid Services 2017). Fairness will also not be addressed by simply using machine learning rather than stepwise regression to estimate the risk adjustment formula or perform variable selection (e.g., Rose 2016; Shrestha et al. 2017) if standard statistical fit metrics are still the basis of evaluation for those tools.

The fairness issues facing plan payment risk adjustment are not entirely unique given the pervasive use of stepwise regressions and similar tools in many applications. Consideration of what makes a formula fair for the lives these health care algorithms touch has thus far been comparatively underdeveloped. Important work has been done in the areas of predictive policing (e.g., Lum and Isaac 2016), recidivism (e.g., Chouldechova 2017), and hospital ratings (e.g., Phillips 2018), for example. Leveraging and adapting these vital advances while expanding fairness approaches for the distinctive needs of health plan payment is a crucial issue moving forward. The Association for Computing Machinery recently issued a statement on automated decision-making describing seven principles for algorithmic transparency and accountability: awareness, access and redress, responsibility, explanation, data provenance, auditability, and validation and testing (Association for Computing Machinery 2017). It will be especially fruitful to bring these principles to bear in a context where the algorithms have a real impact on the welfare of vulnerable groups in the health care system.

Layton, T.J., McGuire, T.G., Van Kleef, R.C., (2016), "Deriving Risk Adjustment Payment Weights to Maximize Efficiency of Health Insurance Markets." NBER Working Paper 22642.

Layton, T.J., Ellis, R.P., McGuire, T.G., and Van Kleef, R.C. (2017), "Measuring Efficiency of Health Plan Payment Systems in Managed Competition Health Insurance Markets," *Journal of Health Economics,* 56, 237-255.

Layton, T.J., Montz, E., and Shepard, M. (2018), "Health Plan Payment in U.S. Marketplaces: Regulated Competition with a Weak Mandate," in McGuire, T.G., and Van Kleef, R.C. (eds.), *Risk Adjustment, Risk Sharing and Premium Regulation in Health Insurance Markets: Theory and Practice*, New York, NY: Elsevier.

Lum, K., and Isaac, W. (2016), "To predict and serve?" *Significance,* 13(5), 14-19.

Mantel, N. (1970), "Why stepdown procedures in variable selection," *Technometrics*, 12, 591—612.

McGuire, T.G. (2016), "Achieving Mental Health Care Parity Might Require Changes in Payment and Competition," *Health Affairs*, 35(6), 1029-1035.

Miller, A.J. (2002), "Subset Selection in Regression," Norwell, MA: CRC Press.

Mitchell, S. (2017). "Fairness: Notation, definitions, data, legality," [online]. Available at https://speak-statistics-to-power.github.io/fairness/old.html.

Montz, E., Layton, T.J., Busch, A., Ellis, R., Rose, S., and McGuire T.G. (2016), "Risk adjustment simulation: Plans may have incentives to distort mental health and substance use coverage," *Health Affairs*, 35(6), 1022–28.

O'Malley, A. J., Zaslavsky, A. M., Elliott, M. N., Zaborski, L., and Cleary, P. D. (2005), "Case-Mix Adjustment of the CAHPS® Hospital Survey," *Health Services Research*, 40, 2162-2181.

Phillips, D. (2018), "At Veterans Hospital in Oregon, a Push for Better Ratings Puts Patients at Risk, Doctors Say," *New York Times* [online]. Available at https://www.nytimes.com/2018/01/01/us/at-veterans-hospital-in-oregon-a-push-for-better-ratings-puts-patients-at-risk-doctors-say.html.

Pope, G.C., Kautter, J., Ellis, R.P., Ash, A.S., Ayanian, J.Z., Ingber, M.J., Levy, J.M., and Robst, J. (2004), "Risk Adjustment Of Medicare Capitation Payments Using The CMS-HCC Model," *Health Care Financing Review*, 25(4), 119-141.

Pope, G. C., Kautter, J., Ingber, J.J., Freeman, S., Sekar, R., and Newhart, C. (2011), "Evaluation of the CMS-HCC Risk Adjustment Model," [online]. Available at http://www.nber.org/risk-adjustment/2011/Evaluation2011/Evaluation_Risk_Adj_Model_2011.pdf

Reyes J. (2016), "Technologists must do better: Drexel prof on the ethics of algorithms." *Technical.ly Philly* [online]. Available at https://technical.ly/philly/2016/09/30/kris-unsworth-ethics-algorithms.
16